\documentclass{ifacconf}

\usepackage{multirow}
\usepackage{amsfonts}

\usepackage{stfloats}

\usepackage{amsmath}
\usepackage{amssymb}
\usepackage{enumerate}
\usepackage{definice}
\usepackage{color}
\usepackage{booktabs,hhline} 
\usepackage{cool}
\usepackage{natbib}
\usepackage{mathtools}

\newcommand{\bbR}{\ensuremath{\mathbb{R}}}

\newcommand{\mcR}{\ensuremath{\mathcal{R}}}

\usepackage[scr=boondoxo,scrscaled=1.05]{mathalfa}

\def\bfD{\mathbf D}

\def\bfF{\mathbf F}
\def\bfG{\mathbf G}

\def\bfI{\mathbf I}

\def\bfL{\mathbf L}
\def\bfM{\mathbf M}

\def\bfP{\mathbf P}

\def\bfS{\mathbf S}
\def\bfT{\mathbf T}

\def\bfZ{\mathbf Z}

\def\bfc{\mathbf c}

\def\bff{\mathbf f}
\def\bfg{\mathbf g}
\def\bfh{\mathbf h}

\def\bfv{\mathbf v}
\def\bfw{\mathbf w}
\def\bfx{\mathbf x}
\def\bfy{\mathbf y}
\def\bfz{\mathbf z}

\def\bfxi{\boldsymbol{\xi}}

\def\bfDelta{\mathbf \Delta}

\def\calO{\mathcal{O}}

\def\calO{\mathbfcal{O}}

\DeclareMathAlphabet\mathbfcal{OMS}{cmsy}{b}{n}

\def\real{\mathbb{R}}

\def\T{^T}

\usepackage{graphicx}      
\usepackage{natbib}        

\def\nx{d}
\def\nu{c}
\def\nz{b}

\newcommand{\tightoverset}[2]{%
	\mathop{#2}\limits^{\vbox to -.5ex{\kern-0.75ex\hbox{$#1$}\vss}}}

\begin{document}
	\begin{frontmatter}
		
		\title{Tensor Train Discrete Grid-Based Filters: Breaking the Curse of Dimensionality  \hspace{-0mm}}
		
				\thanks[footnoteinfo]{The authors would like to express their gratitude to the Dr. Dmitry Savostyanov for the consultation concerning the TT-toolbox, and also to Mr. Andrei Chertkov for his advice. This work was co-funded by the European Union under the project ROBOPROX- Robotics and advanced industrial production(reg.no.CZ.02.01.01/00/22\_008/0004590).}
		
\author[]{J. Matou\v{s}ek$^*$, M. Brandner$^\dagger$,  J. Dun\'{i}k$^*$, I. Pun\v{c}och\'{a}\v{r}$^*$}

\address[]{Dept. of Cybernetics, $^\dagger$Dept. of Mathematics \\University of West Bohemia, Pilsen, Czech Republic\\ e-mails: \{matoujak, dunikj, ivop\}@kky.zcu.cz, brandner@kma.zcu.cz}

		\begin{abstract} 
		This paper deals with the state estimation of stochastic systems and examines the possible employment of tensor decompositions in grid-based filtering routines, in particular, the tensor-train decomposition. The aim is to show that these techniques can lead to a massive reduction in both the computational and storage complexity of grid-based filtering algorithms without considerable tradeoffs in accuracy. This claim is supported by an algorithm descriptions and numerical illustrations.\footnote{\copyright 2024 the authors. This work has been accepted to IFAC for publication under a Creative Commons Licence CC-BY-NC-ND.}
		\end{abstract}
		
		\begin{keyword}
			State estimation, tensor decomposition, tensor-train, point-mass method
		\end{keyword}
		
	\end{frontmatter}

	\section{Introduction}
The problem of recursive state estimation of nonlinear discrete-time stochastic dynamic systems from noisy measured data has been a subject of considerable research interest for the last several decades. In this paper, the discrete-time state-space model of a nonlinear stochastic dynamic system with additive noises
\begin{align}
\bfx_{k+1}&=\bff_{k}(\bfx_{k})+\bfw_{k}, \label{eq:asx}\\
\bfz_{k}&=\bfh_k(\bfx_k)+\bfv_k,\label{eq:asz}
\end{align}
is considered. The vectors $\bfx_k\in\real^{\nx}$, and $\bfz_k\in\real^{\nz}$ represent the \textit{unknown} 
state of the system, and the \textit{known} realization of a measurement at time instant $k$, respectively. The state and measurement functions $\bff_k:\real^{\nx}\rightarrow\real^{\nx}$ and $\bfh_k:\real^{\nx}\rightarrow\real^{\nz}$ are supposed to be \textit{known}. Particular realizations of the state and measurement noises $\bfw_k$ and $\bfv_k$ are \textit{unknown}, but their probability density functions (PDFs) are known. The state noise random variable (RV) $\bfw_k$ and the measurement noise RV $\bfv_k$ are mutually independent, and independent of the \textit{known} initial state RV $\bfx_0$. 

Bayesian state estimation calculates a PDF of the state $\bfx_k$ conditioned on all measurements $\bfz^\ell=[\bfz_0,\bfz_1,\ldots,\bfz_\ell]$  up to the time instant $\ell$, i.e. the \textit{conditional} PDF $p(\bfx_k|\bfz^\ell)$, is sought. The general solution to the state estimation is represented by the Bayesian recursive relations (BRRs) which describe the evolution of the PDFs \citep{AnMo:79}, namely by the Chapman-Kolmogorov equation (CKE) for the predictive PDF $p(\bfx_{k}|\bfz^{k-1})$ and the Bayes' rule for the filtering PDF $p(\bfx_{k}|\bfz^k)$ calculation. Unfortunately, the analytical solution to the BRRs is intractable for a vast majority of practical systems and the approximate solution is sought.

In this paper, we focus on approximate numerical solution to the BRRs using the highly accurate grid-based filters (GBFs) \citep{SiKraSo:06}, which are, however, hindered by the curse of dimensionality. The curse of dimensionality reduction without introducing additional approximations or assumptions, has been treated only marginally in the literature. In this area, we can mention the grid-based state estimation of continuous and discrete state-space models using a sparse grid design and application of the discrete  fast Fourier transform (FFT), respectively. Unfortunately, any of these methods cannot be easily and exactly extended for nonlinear discrete dynamics, either due to ambiguity of the grid cells sizes  \citep{KaSch13} or non-uniform predictive grid \citep{MaDuBrPaCh:23}. 

Therefore, we aim on different direction of research on computational complexity (CC) reduction, which is application of tensor-train decomposition (TTD). First, we utilise the TTD to reduce CC and storage requirements, which are exponentially growing with dimension. Second, we employ the FFT to reduce CC scaling with number of points ``discretising'' the state space.

	\section{Grid-Based Estimation}
	Numerical grid-based solution to BRRs is based on an approximation of a PDF $p(\bfx_k)$ by a \textit{piece-wise constant} point-mass density (PMD) computed at the set of $N$ grid points $\bfx_k^{(:)}=\{\bfx_k^{(i)}\}_{i=1}^N , \bfx^{(i)}_k\in\real^{d}$, which are in the middle of their respective non-overlapping neighbourhood $\bfDelta_k^{(i)}$, as \citep{SiKraSo:06,DuMaSt:23}
	\begin{align}
	p(\bfx_k;\bfx_k^{(:)})\triangleq\sum_{i=1}^N\bfP_{k}^{(i)}S(\bfx_k;\bfDelta_k^{(i)}),\label{eq:pdf_pm}
	\end{align}
	where $N = N_1 \cdot N_2\ ... \cdot N_{d}$, and $N_i$ is a number\footnote{If $N_1 = N_2 = ... = N_{d} = N_{\mathrm{pa}}$, then  the notation $N_{pa}$ is used.} of discretisation points in $i$-th dimension of the state $\bfx_k$, $P_{k}^{(i)} = p(\bfx_k^{(i)})$ is the value of the PDF $p(\bfx_k)$ evaluated at the $i$-th grid point $\bfx^{(i)}_k$ further also called as a \textit{weight} (the PMD is normalized to integrate to 1), $\bfDelta_k^{(i)}$ is rectangular grid cell (i.e. the grid is equidistant) centred at $\bfx^{(i)}_k\in\real^{d}$ with the volume $\delta_k$ , where $p(\bfx_k;\bfx_k^{(:)})$ is constant, and $S(\bfx_k;\bfDelta_k^{(i)})$ is an indicator function that equals to 1 if $\!\bfx_k \in \bfDelta_k^{(i)}$.


%
	
	\subsection{Time Update in Vectorised and Tensor Notation}
	Let the filtering PMD be 
	\begin{align}
	p(\bfx_k|\bfz^k;\bfx_k^{(:)})\triangleq\sum_{i=1}^N\bfP_{k|k}^{(i)}S(\bfx_k;\bfDelta^{(i)}_k)\label{eq:filtPMD}
	\end{align}	
	and the predictive or new grid $\bfx_{k+1}^{(:)}=\{\bfx_{k+1}^{(i)}\}_{j=1}^N$, is covering the suitable part of the state-space for time $(k+1)$. The predictive grid $\bfx_{k+1}^{(:)}$ can be obtained using \textit{(i)} a transformation of the ``old'' grid $\bfx_{k}^{(:)}$ via the dynamics \eqref{eq:asx} and subsequent rectangularisation or \textit{(ii)} moment-based prediction \citep{DuMaSt:23}. Then, the numerical solution to the CKE leads to
	\begin{align}
	p(\bfx_{k+1}|\bfz^k;\bfx_{k+1}^{(:)})=\sum_{j=1}^N \bfP_{{k+1|k}}^{(j)}S(\bfx_{k+1};\bfDelta_{k+1}^{(j)}).\label{eq:pdf_pm_pred}
	\end{align} 
	\subsubsection{Vectorised Form:}
	In the standard matrix time-update solution, the value of the predictive PDF at $j$-th grid point is computed \citep{SiKraSo:06}
	\begin{align}
	\bfP_{{k+1|k}}^{(j)}\!=\!\sum_{i=1}^N p(\bfx^{(j)}_{k+1}|\bfx^{(i)}_{k})\bfP_{k|k}^{(i)}\delta_k,\label{eq:pdf_predII}
	\end{align} 
	which can be conveniently written in a \textit{matrix} form as
	\begin{align}
	\bfP_{{k+1|k}}^{(:)} =\bfT^\mathrm{m} \bfP_{{k|k}}^{(:)}. \label{eq:CKEnumSol}
	\end{align}
	In \eqref{eq:CKEnumSol}, $\bfP_{{k}}^{(:)}$ denotes all weights stacked in a vector, and $\bfT^\mathrm{m}\in\real^{N\times N}$ is the transition probability matrix (TPM), with the element in $j$-th row and $i$-th column given by 
	\begin{align}
	\bfT^{\mathrm{m}}_{{j,i}} &= p(\bfx^{(j)}_{k+1}|\bfx^{(i)}_k)\delta_k. \label{eq:TPMmat}
	\end{align}

\begin{figure*}[t]
	\centering 
	\includegraphics[width=0.7\linewidth]{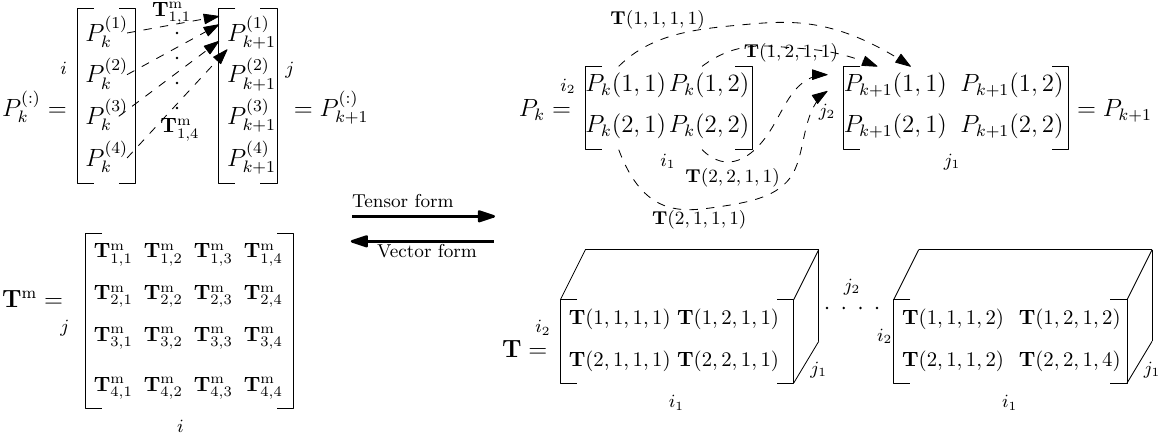}
	\vspace{-0.5cm}\caption{Illustration of vectorized time-update and time-update in tensor form.}
	\label{fig:TensorIllust}
   \end{figure*}

	The reader might notice that this standard formulation \eqref{eq:pdf_predII}--\eqref{eq:TPMmat} is, in fact, a vectorized form. It means, that the vectors of weights $\bfP_{{k}}^{(:)}, 	\bfP_{{k+1}}^{(:)} $ comes generally from tensors $\bfP_{{k}}, \bfP_{{k+1}} \in \mathbb{R}^{N_1 \times N_2 ... \times N_d}$ as well as the TPM  $\bfT^\mathrm{m}$ is a reshaped tensor $ \bfT \in \mathbb{R}^{N_1 \times N_2 ... \times N_d \times N_1 \times N_2 ... \times N_d}$. Transformation between matrix and tensor form is illustrated in Fig. \ref{fig:TensorIllust} for $d = 2$.

	\subsubsection{Tensor Form:}
	
As the standard calculation using the TPM suffers from the curse of dimensionality (term \eqref{eq:TPMmat} needs to be evaluated $N^2=(N_{\mathrm{pa}}^d)^2$ for all combination of the new and old grid points), the efficient version was employed \citep{MaDuBr:23, MaDuBrPaCh:23, DuMaSt:23}. It was shown that under conditions of odd number of points per dimension and new grid points calculation as $\bfx_{k+1}^{(:)} = \bff(\bfx_{k}^{(:)})$, the predictive weights $ \bfP_{{k+1}}$ can be calculated efficiently by the FFT-based convolution as
 \begin{align}
		\bfP_{{k+1|k}}= \mathcal{F}^{-1}\left(  \mathcal{F}({\bfT}^{\mathrm{mid}}) \odot \mathcal{F}({\bfP}_{{k|k}})\right), \label{eq:tpmconvFFT}
		\end{align}
where $\mathcal{F}$ denotes the discrete FFT, $\odot$ the Hadamard product, and ${\bfT}^{\mathrm{mid}} \in \mathbb{R}^{N_1, N_2, ... N_d}$ a tensor constructed from the middle row of the TPM $\bfT^\mathrm{m}$. This reduces the CC from $\calO(N^2)$ to $\calO(N\log N)$. This efficient prediction inherently leads to the TPM and PMD in the tensor form.


	\subsection{Measurement Update}
	
Independently of the PMD formulation (vector or tensor form), the filtering PMD that uses measurement $\bfz_{k+1}$ is
\begin{align}
p(\bfx_{k+1}|\textbf{z}^{k+1};\bfx_{k+1}^{(:)})
=\sum_{i=1}^N \bfP_{k+1|k+1}^{(i)}S(\bfx_{k+1};\bfDelta_{k+1}^{(i)}),\label{eq:pmf_filt_pfiltII}
\end{align}
where the weight at $\bfx_{k+1}^{(i)}$ is
\begin{align}
\bfP_{k+1|k+1}^{(i)}&={\tilde{c}}_{k+1}^{-1}p(\textbf{z}_{k+1}|\bfx_{k+1}=\bfx_{k+1}^{(i)})\bfP_{k+1|k}^{(i)}\label{eq:pmf_filtrace_a}.
\end{align}
The normalization constant equals ${\tilde{c}}_{k+1}$
\begin{align}
\tilde{c}_{k+1}&=\sum_{i=1}^N p(\bfz_{k+1}|\bfx_{k+1}=\bfx_{k+1}^{(i)})\bfP_{k+1|k}^{(i)}\delta_{k+1}\label{eq:pmf_filtrace_b}.
\end{align} 
Considering the vectorised notation \footnote{Tensor formulation can be seen further.}, the measurement step \eqref{eq:pmf_filtrace_a} becomes
\begin{align}
	\bfP_{k|k}^{(:)} \propto \bfL^\mathrm{v} \odot \bfP_{k}^{(:)},
\end{align}
where $\bfL^\mathrm{v}$ is a vector of likelihoods $p(\textbf{z}_{k+1}|\bfx_{k+1}=\bfx_{k+1}^{(i)})$. 

\subsection{Limitation and Motivation}
There are two limiting properties of the standard, i.e., vectorised, grid-based filter; \textit{(i)} CC of the prediction step (growing \textit{quadratically} with $N$), 
\textit{(ii)} total number of grid points (growing \textit{exponentially} with $d$). Whereas the former limitation was solved by application of the FFT in the tensor framework, the latter has not been addressed yet.

In this paper, we propose to take advantage of the tensor notation and to decompose the considered tensors (transition probability, conditional PMDs, and the likelihood) using the \textit{tensor train decomposition} (TTD) discussed e.g. in \citep{OsTy:10}. The TTD is a powerful tool to express high-dimensional tensors with reduced storage complexity, which is then enhanced with the FFT to further reduce the CC.

	\section{Tensor Train Decomposition}
Tensors are multidimensional generalization of matrices. As shown earlier, they naturally arise in grid based state estimation for a higher dimensional state. Let $A \in \mathbb{R}^{N_1 \times N_2 ... \times N_d} $ be a $d$-dimensional tensor with elements
\begin{align}
	 A(i_1,i_2,...,i_d).
\end{align}
\subsection{Tensor Decompositions}
The tensor $A$ can be represented in a decomposed form limiting the number of stored values. The widely used decompositions are \citep{OsTy:10}:
\begin{enumerate}
	\item CANDECOMP decomposition: 
	\begin{align}
		\begin{smallmatrix}
			A(i_1,i_2,...,i_d) = \sum_{r=1}^\mathcal{R} U_1(i_1,r)U_2(i_2,r)\cdots U_d(i_d,r),
		\end{smallmatrix}
	\end{align}
	where $U_l \in \mathbb{R}^{N_\ell \times \mathcal{R}}$, $N_\ell$ is the number of elements in dimension $\ell$. Its limitation lies mainly in missing robust/fast numerical methods for reduction of rank $\mathcal{R}$.
	\item Tucker decomposition :
	\begin{align}
		\begin{smallmatrix}
			A(i_1,i_2,...,i_d) = \sum^{\mathcal{R}_1, \mathcal{R}_2 \cdots \mathcal{R}_d}_{r_1,r_2, \cdots r_d} {C}(r_1,r_2, \cdots r_d) U_1(i_1,r_1)\\\cdot U_2(i_2,r_1)\cdots U_d(i_d,r_1),
		\end{smallmatrix}
	\end{align} 
	with Tucker ranks $\mathcal{R}_1, \mathcal{R}_2 \cdots \mathcal{R}_d$, is suitable for lower dimensions because ``multi-dimensional'' tensor core $C  \in \mathbb{R}^{r_1 \times r_2 \cdots \times r_d}$ has to be stored.
	\item TTD \citep{Os:11}: The name stems from the tensor cores being connected by one index forming in a sense a train of three dimensional tensors. The decomposition is 
	\begin{align}
		\begin{smallmatrix}
			A(i_1,i_2,...,i_d) =  \sum_{r_0...r_d}^{\mathcal{R}_1\cdots \mathcal{R}_d} G_1(r_0,i_1,r_1)G_2(r_1,i_2,r_2)\cdots \\ \cdot G_d(r_{d-1},i_d,r_d).
		\end{smallmatrix}
		 \label{eq:ttd_exact}
	\end{align}
	To ensure that the product is a scalar, the first and last rank is  $\mathcal{R}_0 = \mathcal{R}_d = 1$. The main advantage of the TT lies in efficient and effective recompression, i.e., the ranks increased by a numerical operation can be easily decreased.
\end{enumerate}
Based on the brief description of the properties of various tensor decompositions together with decomposition storage complexity summarised in Table \ref{tab:compCplx}, the TTD is currently regarded as the \textit{most efficient} decomposition and, thus, it is further considered in this paper in design of the grid-based filter.

\begin{table}[]
\begin{tabular}{lll}
 CADENCOMP                           & Tucker                                                          & Tensor Train                                 \\ \hline
 $N_{pa} d \mathcal{R}$ & $N_{pa} d \max(\mathcal{R}) + \max(\mathcal{R})^d$ & $N_{pa} d \max(\mathcal{R})^2$
\end{tabular}\vspace*{2mm}\caption{Storage complexity of decompositions \citep{Ci:14}}
\label{tab:compCplx}
\end{table}

\subsection{High-dimensional Tensor: Accuracy vs. Memory}

All decompositions, TTD including, are exact for high enough rank $\mathcal{R}$ or ranks $\mathcal{R}_1, \mathcal{R}_2, \ldots, \mathcal{R}_d$. However for a high dimension $d$, a tensor cannot be explicitly saved in the computer memory (no matter if decomposed), and it is even impossible to perform any arbitrary mathematical operation on each of its elements. Therefore, it is essential to develop and employ an approximation allowing the manipulation with high dimensional tensors. This approximation is usually called \textit{low-rank-tensor} approximation, and it imposes a restriction on the maximum allowed rank $\mathcal{R^\mathrm{max}}$ at the cost of  an \textit{approximate}  decomposition. 

\subsection{Black-Box Decomposition Approximation}
In this paper, we adopts the so called \textit{black-box} decomposition approximation \citep{GrKrTo:13}, where the only requirement is the ability to evaluate the tensor at an arbitrary position. 
 These algorithms are usually based on a matrix approximative decompositions (such asi SVD).

  In our case the base of the black-box approximation is an efficient approximative cross/skeleton decomposition. For $d = 2$, this decomposition is given by \citep{GoTyZa:97}
\begin{align}
	A \approx C \hat{A}^{-1} R,\label{eq:matrixDecomp}
\end{align}
where $R$ is a matrix of $\mathcal{R}$ rows of $A$, $C$ is a matrix of $\mathcal{R}$ columns of $A$, and $\hat{A}$ is a sub-matrix of elements that are on the cross, i.e. where the chosen rows and columns intersect. The accuracy of the approximation depends on the rows and columns selected. For given $\mathcal{R}$ it was shown \citep{GoTy:01} that a quasi-optimal approximation is yielded when the rows and columns are chosen so that the cross determinant $\det(\hat{A})$ is the largest possible. However, the problem of finding the maximum determinant is NP hard. Therefore, realistically some heuristic algorithm has to be used to find a sub-optimal cross. These algorithms, called \textit{cross approximation} algorithms, are briefly overviewed in \citep{Sa:14}.

Out of these sub-optimal cross interpolation algorithms, we adopted the \textit{greedy restricted cross interpolation algorithm} for tensor trains, where the cross selection is performed by random selection of the rows and columns  \citep{Sa:14}. The greedy interpolation, based on the matrix decompositions \eqref{eq:matrixDecomp}, is implemented in the MATLAB\textregistered\ TT-toolbox \citep{TTtoolbox}. 

\section{Design of Tensor Train Grid-Based Filters}
Having introduced concept of the grid-based filter in tensor format (Section 2) and TTD (Section 3), the TT grid-based filter can be developed for both \textit{(i)} standard, and \textit{(ii)} FFT-based formulations. Developed filtering algorithms take advantage of the following notation:
\begin{itemize}
	\item The grid is chosen to be equidistantly spaced and can be therefore stored axis per axis using only $N_S = N_1 + N_2 +\ldots + N_d$ values as opposed to the $N = N_1  N_2 \ldots  N_d$ values for an arbitrary grid.
	\item Instead of vectorized weights $\bfP_{{k}}^{(:)}$, the weights are stored in the tensor format corresponding to the underlying grid shape, with elements $\bfP_k(i_1,i_2,...,i_d)$. 
	\item Consequently, the TPM $ \bfT^\mathrm{m} \in \mathbb{R}^{N \times N}$ is further a tensor $ \bfT \in \mathbb{R}^{N_1 \times N_2 ... \times N_d \times N_1 \times N_2 ... \times N_d}$ and the prediction is performed as $\bfT^{i_1 i_2...i_{2d}} \bfP_{i_{d+1} i_{d+2} ... i_{2d}}$ written in an Einstein summation notation.
	\item The vector of likelihood weights $\bfL^\mathrm{v}$ is treated as a tensor $ \bfL \in \mathbb{R}^{N_1 \times N_2 ... \times N_d }$.
\end{itemize}

Efficient design of the tensor train (TT) grid-based filter requires a set of function for TT approximation and manipulation. Most of the functions are part of the TT-toolbox, the remaining have been specifically developed.

\subsection{Toolbox Functions}

Some function that were needed are part of the TT-toolbox, these are 
\begin{itemize}
	\item Cross approximation of the tensors in TT format, routine \textit{greedy2\_cross}.
	\item Dot product of two tensors in a tensor train format, routine \textit{dot}.
	\item Hadamard product of two tensors \textit{times}.
	\item The Hadamard product in the TT format is the Kronecker product of the cores $G$, which essentially squares the rank sizes of the core. Then, a rounding routine is needed to reduce the ranks  to the required accuracy of the TT approximation, routine \textit{round}.
	\item For normalization of the densities a division of a TT by a number is needed, routine \textit{mrdivide}, and also summation of all elements, routine  \textit{sum}.
	\item Also functions that transform the tt to its cores and back, routines \textit{cell2core}, \textit{core2cell}.
	\item Power for mean calculation, routine \textit{power}.
\end{itemize}

\subsection{Developed Functions}
This section presents functions that had to be implemented in MATLAB\textregistered, as they were not part of the TT toolbox. The Einstein summation algorithm was developed from the scratch. Remaining algorithms were taken from the literature.

\subsubsection{Einstein Summation}

The Einstein summation algorithm for the TT format was inspired by \citep{KiCaKoLeMa:21}, where the summation is presented for single index only. 

Let us assume that the transition probability tensor $\bfT\in\mathbb{R}^{N_{1}\times \ldots\times N_{d} \times N_{1} \times \ldots \times N_{d}}$ has the following tensor train decomposition
\begin{multline}
	\begin{smallmatrix}
		\bfT(i_{1},i_{2},\ldots,i_{d},i_{d+1},\ldots,i_{2d}) =
		\sum_{r_{0}=1}^{\mcR_{0}}\ldots\sum_{r_{2d}=1}^{\mcR_{2d}}
		\bfT_{1}(r_{0},i_{1},r_{1})\ldots\\\bfT_{2d}(r_{2d-1},i_{2d},r_{2d}),
	\end{smallmatrix}
\end{multline}
where the sums over outer ranks $\mcR_{0}=\mcR_{2d}=1$ are written explicitly to have a uniform notation in the algorithm. Similarly the tensor of weights\footnote{The time subscript $k$ is omitted here to simplify notation.} $\bfP\in\bbR^{N_{1}\times\ldots\times N_{d}}$ has the tensor train decomposition
\begin{multline}
	\begin{smallmatrix}
		\bfP(i_{d+1},\ldots,i_{2d}) = \sum_{r_{d}'=1}^{\mcR_{d}'}\ldots\sum_{r_{2d}'=1}^{\mcR_{2d}'}
		\bfP_{d+1}(r_{d}',i_{d+1},r_{d+1}')\ldots\\\bfP_{2d}(r_{2d-1}',i_{2d},r_{2d}'),
	\end{smallmatrix}\hspace*{3mm}
\end{multline}
where the indices are intentionally shifted by $d$ to have alignment with the tensor train decomposition of $\bfT$ and the sum over outer ranks $\mcR_{d}'=\mcR_{2d}'=1$ are again written explicitly. The tensor of predictive weights for the next time step $\bfP'\in\bbR^{N_{1}\times\ldots\times N_{d}}$ is calculated as
\begin{multline}\label{ali:ttConvolutionP}
	\begin{smallmatrix}
		\bfP'(i_{1},\ldots,i_{d}) = \sum_{i_{d+1}=1}^{N_{1}} \ldots \sum_{i_{2d}=1}^{N_{d}}
		\bfT(i_{1},i_{2},\ldots,i_{d},i_{d+1},\ldots,i_{2d})\\ \cdot\bfP(i_{d+1},\ldots,i_{2d}).
	\end{smallmatrix}\hspace*{1mm}
\end{multline}
If we substitute the tensor train decompositions and change the order of summation, we can write $\bfP'$ as
\begin{multline}
	\begin{smallmatrix}
		\bfP'(i_{1},\ldots,i_{d}) = \sum_{r_{0}=1}^{\mcR_{0}} \ldots \sum_{r_{d-1}=1}^{\mcR_{d-1}}\sum_{r_{d}'=1}^{\mcR_{d}'}
		\bfT_{1}(r_{0},i_{1},r_{1}) \ldots \\
		\bfT_{d-1}(r_{d-2},i_{d-1},r_{d-1}) 
		\cdot\sum_{r_{d}=1}^{\mcR_{d}} \bfT_{d}(r_{d-1},i_{d},r_{d})\bfZ(r_{d},r_{d}'),
	\end{smallmatrix}\hspace*{4mm}
\end{multline}
where $\bfZ\in\bbR^{\mcR_{d}\times \mcR_{d}'}$ is a one-dimensional tensor ($\mcR_{d}'=1$) given as
\begin{multline} \label{ali:ttConvolutionZ}
	\begin{smallmatrix}
		\bfZ(r_{d},r_{d}') = \sum_{r_{d+1}=1}^{\mcR_{d+1}}\ldots\sum_{r_{2d}=1}^{\mcR_{2d}} \sum_{r_{d+1}'=1}^{\mcR_{d+1}'} \ldots \sum_{r_{2d}'=1}^{\mcR_{2d}'}\\
		\bfM_{d+1}(r_{d},r_{d+1},r_{d}',r_{d+1}') \ldots \bfM_{2d}(r_{2d-1},r_{2d},r_{2d-1}',r_{2d}').
	\end{smallmatrix}\hspace*{8mm}
\end{multline}
For each $k=d+1,d+2,\ldots,2d$, the auxiliary four-dimensional tensor $\bfM_{k}\in\bbR^{\mcR_{k-1}\times\mcR_{k}\times\mcR_{k-1}'\times\mcR_{k}'}$ is given as
\begin{multline}\label{ali:ttConvolutionM}
	\begin{smallmatrix}
		\bfM_{k}(r_{k-1},r_{k},r_{k-1}',r_{k}') =
		\sum_{i_{k}=1}^{N_{k}} \bfT_{k}(r_{k-1},i_{k},r_{k})\bfP_{k}(r_{k-1}',i_{k},r_{k}').
	\end{smallmatrix}\hspace*{-4mm}
\end{multline}
Note that the tensor $\bfZ$ can be computed recursively by going forward or backward through dimensions. The backward recursion can be written as follows
\begin{multline}\label{ali:ttConvolutionZrecursion}
	\begin{smallmatrix}
		\bfZ^{(j)}(r_{j},r_{j}') = \sum_{r_{j+1}=1}^{\mcR_{j+1}}\sum_{r_{j+1}'=1}^{\mcR_{j+1}'} \bfM_{j+1}(r_{j},r_{j+1},r_{j}',r_{j+1}')\cdot\\
		\bfZ^{(j+1)}(r_{j+1},r_{j+1}'),
	\end{smallmatrix}\hspace*{3mm}
\end{multline}
where $\bfZ^{(j)}\in\bbR^{\mcR_{j}\times \mcR_{j}'}$ is a two-dimensional tensor, $\bfZ^{(2d)}=1$ is the initial value and $j=2d-1,2d-2,\ldots,d$ is an index that goes backward through dimensions. After the recursive computation is finished, we have $\bfZ(r_{d},r_{d}')=\bfZ^{(d)}(r_{d},r_{d}')$.

The operations with the tensors in~\eqref{ali:ttConvolutionP}, \eqref{ali:ttConvolutionZrecursion}, and \eqref{ali:ttConvolutionM} can be performed using matrix multiplication when suitable permutation of indices and reshaping of tensors is performed. A MATLAB\textregistered\  pseudo-code is summarized in Algorithm 1.

\noindent\rule{8.5cm}{0.5pt}
\newline
\textbf{Algorithm 1: Einstein Summation}
\vspace*{-1mm}
\begin{enumerate}
	\setcounter{enumi}{0}
	\item \textit{Initialization:} Set a temporary variable to scalar value $\bfZ_{\mathrm{mat}}=1$ and an index $j=d$.
	\item \textit{Create matrix representations of tensor cores $\bfT_{j+d}$ and $\bfP_{j}$:} Compute $\bfT_{\mathrm{mat}} = \mathtt{reshape}(\mathtt{permute}(\bfT_{j+d},\allowbreak [1,3,2]),\mcR_{j+d-1}\mcR_{j+d},N_{j+d})$,\\ $\bfP_{\mathrm{mat}} = \mathtt{reshape}(\mathtt{permute}(\bfP_{j},\allowbreak[1,3,2]),\mcR_{j-1}'\mcR_{j}',N_{j})$.
	\item \textit{Create matrix representation of tensor $\bfM$:} Compute $\bfM_{\mathrm{mat}} = \mathtt{reshape}(\mathtt{permute}(\mathtt{reshape}(\bfT_{\mathrm{mat}}\bfP_{\mathrm{mat}}^{\T},\allowbreak \mcR_{j+d-1},\mcR_{j+d},\mcR_{j-1}',\mcR_{j}'),[1\ 3\ 2\ 4]),\\ \mcR_{j+d-1}\mcR_{j-1}', \mcR_{j+d}\mcR_{j}')$
	\item \textit{Update $\bfZ_{\mathrm{mat}}$}: Compute $\bfZ_{\mathrm{mat}} = \bfM_{\mathrm{mat}}\bfZ_{\mathrm{mat}}$.
	\item \textit{Set $j=j-1$. If $j>0$ go to step (3).}
	\item \textit{Compute tensor cores $\bfP_{j}'$ for $j=1,\ldots,d-1$:} Set $\bfP_{j}'=\bfT_{j}$ for  $j=1,\ldots,d-1$.
	\item \textit{Compute tensor core $\bfP'_{d}$:} Compute\\ $\bfP'_{d}=\mathtt{reshape}(\mathtt{reshape}(\bfT_{d},\mcR_{d-1}N_{d},\mcR_{d})\bfZ_{\mathrm{mat}},\\ \mcR_{d-1},N_{d})$
\end{enumerate}
\vspace*{-3mm}
\rule{8.5cm}{0.5pt}
\newline
\vspace*{-3mm}

\subsubsection{FFT Convolution in Tensor Train}

FFT based convolution in TT format was proposed in \citep{RaOs:16} and it was implemented without any modifications. It was shown there, that the tensor cores resulting from the convolution of two tensors can be calculated simply as the inverse FFT of Hadamard product of FFTs of their cores.

\subsubsection{Interpolation in Tensor Train}

Interpolation in TT format was proposed in \citep{OlRyCo:17} and it was implemented on the level of the decomposed cores.

\subsection{Tensor Train Grid-based Filters}
In this section the \textit{proposed} algorithms of the grid-based filters in TT format are presented in a compact form. 

\subsubsection{Standard Filter in TT Format}
First, the standard formulation of the filter in TT format is presented in Algorithm 2.
\newline
\rule{8.5cm}{0.5pt}
\newline
\textbf{Algorithm 2: Point-Mass Filter in Tensor Train Format}
\vspace*{-1mm}
\begin{enumerate}
	\item \textit{Initialisation}: Set $k=0$, construct the initial grid of points, that is define vectors of grid points coordinates per axis $\bfxi_0^1, \bfxi_0^2 ... \bfxi_0^d$ and define the tensor train of the weights of initial PMD $\bfP_0$ (\textit{greedy2cross}).
	
	\item  \textit{Meas. update}: Construct the likelihood tensor train $\bfL$ (\textit{greedy2cross}), and calculate the filtering PMD weights as $\bfP_{k|k} = \bfL \odot \bfP_{k-1}$ (\textit{times}). Normalize the result to prevent numerical problems (\textit{sum, mrdivide}) Round the tensor train to required precision (\textit{round}),

	\item \textit{Grid construction:} Calculate the expected mean and covariance of the predictive estimate (\textit{power, dot}) by a local filter and form a grid around expected mean with size based on expected covariance.
	\item \textit{Time update}: Construct the TPM tensor $\bfT$ (\textit{greedy2cross}). Perform the time update as an Einstein summation $ \bfP_{k+1;i_{1} i_{2} ... i_{d}} = \bfT^{i_1 i_2...i_{2d}} \bfP_{k;i_{d+1} i_{d+2} ... i_{2d}}$. Normalize and round the result (\textit{sum, mrdivide, round, core2cell, cell2core}).
	\item Set $k=k+1$ and go to step 2).
\end{enumerate}
\vspace*{-3mm}
\rule{8.5cm}{0.5pt}
\newline
\vspace*{-3mm}

\subsubsection{FFT-based Filter in TT Format}
This algorithm is based on the FFT version of the grid-based filter described earlier \citep{RaOs:16}. In this implementation described by Algorithms 3 and 4, the interpolation in the TT has to be used twice  to keep the grid boundaries aligned with the state space axes, leading to an efficient storage of the grid. The model dynamics is assumed to be linear, i.e.,  $\bfF \bfx_k$ instead of $\bff(\bfx_k)$, which stems from the FFT filter formulation \citep{MaDuBrPaCh:23}.

\noindent\rule{8.5cm}{0.5pt}
\newline
\textbf{Algorithm 3: Efficient Point-Mass Filter in Tensor Train Format}
\vspace*{-1mm}
\begin{enumerate}\setcounter{enumi}{0}
	\item \textit{Initialisation}: Same as algorithm 2.
	
	\item  \textit{Meas. update}: Same as algorithm 2.
	
	\item \textit{Grid re-design:} See Algorithm 4. 

		\item \textit{Time update}: Compute the temporary predictive weights $\bfP^t_{{k+1}}= \mathcal{F}^{-1}\left(  \mathcal{F}({\bfT}^{l}) \odot \mathcal{F}({\hat{\bfP}}_{{k|k}})\right)$ 
		\item \textit{Grid re-design:} Because the calculated predicted weights hold on a grid that does not have the boundaries aligned with axes, take the corners of the predictive and similarly to the steps 4 and 5 in Algorithm 4, design as small grid as possible circumscribing all of the corners and interpolate the calculated temporary predictive weights $\bfP^{t}_{k+1}$ to the new efficient grid $\bfP_{k+1}$.	
		\item Set $k=k+1$ and go to step 2).
\end{enumerate}
\vspace*{-3mm}
\rule{8.5cm}{0.5pt}
\newline
\vspace*{-3mm}

\noindent\rule{8.5cm}{0.5pt}
\newline
\textbf{Algorithm 4: Grid Re-design}
\vspace*{-1mm}
\begin{enumerate}
\setcounter{enumi}{0}
	\item \textit{Approximate predictive moments}: The first two predictive moments $\hat{\bfx}_{k+1|k}$, ${P}_{k+1|k}$ (\textit{power, dot}) are calculated by a local filter (extended, unscented Kalman filter).
	\item \textit{Set corners of required predictive grid $\bfc_i^{\text{pred}}$:} The grid is based on the predictive moments and Chebyshev inequality/$\sigma$ ellipse probability. Therefore its center is at $\hat{\bfx}_{k+1|k}$, and size is based on ${P}_{k+1|k}$.
	\item \textit{Transform the corners to filtration density space:} $\bfc_i^{\text{meas}} = \bfF^{-1} \bfc_i^{\text{pred}}$.
	\item \textit{Design the new filtering grid $\bfxi_{k}^{\text{new;1,...,d}}$:} It is designed, so that it is as small as possible while circumscribing all corners $\bfc_i^{\text{meas}}$, and having boundaries aligned with state-space axes.
	\item \textit{Interpolate the filtering density weights $\bfP_{k|k}$ on $\bfxi_{k}^{\text{new;1,...,d}}$ forming $\hat{\bfP}_{k|k}$} (\textit{greedy2cross}). 
\end{enumerate}
\vspace*{-3mm}
\rule{8.5cm}{0.5pt}
\newline
\vspace*{-3mm}

\section{Numerical Verification}

	\begin{figure}[t]
	\centering 
	\includegraphics[width=0.9\linewidth]{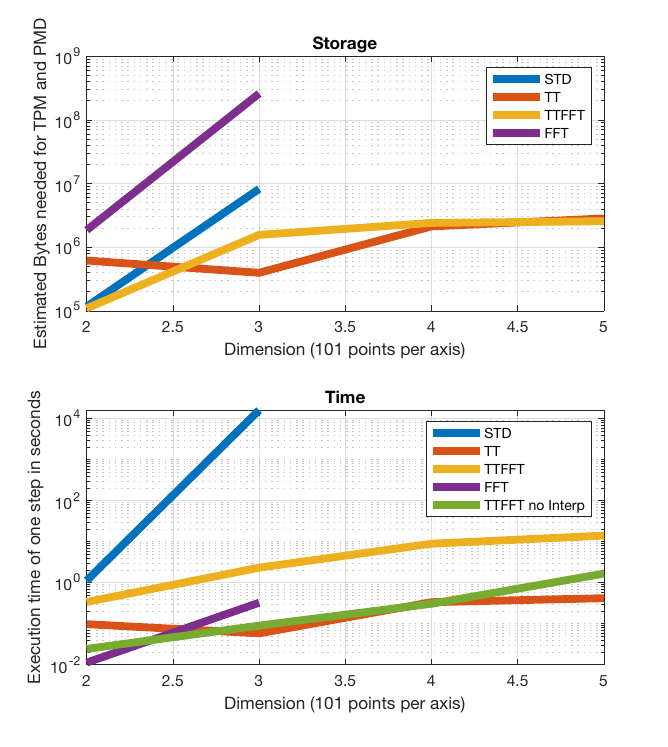}
	\vspace{-0.5cm}\caption{Storage requirements and execution times.}
	\label{fig:timeStor}
   \end{figure}

	A motion model with $n_x=2$ was used for the numerical evaluation. The state  $\bfx_k = \left[ {x_k}, {y_k}\right]$ consisting of horizontal object position evolving according to
		\begin{align}
			\bfx_{k+1} = \begin{bmatrix}
				1.1 & 0.1 \\
				-0.2 & 1.1
			\end{bmatrix} \bfx_{k} + \bfw_k, k = 1,...10
	\end{align}
	where the end simulation time is $k_\mathrm{f} = 10$, and the state noise 	is described by the Gaussian PDF $p(\bfw_k)=\mathcal{N}\{\bfw_k; [0\ 0]^T,\bfI_2\}$. A radar measures  the angle and the range of the object w.r.t. the origin
	\begin{align}
		\bfz_{k} = \left[\begin{smallmatrix}
			\sqrt{(x_k^2 + y_k^2)} \\
			\arctan(y_k, x_k)
		\end{smallmatrix}\right] + \bfv_k,
	\end{align}
	where $\arctan$ is the four quadrant arctangent in [\textit{deg}] and measurement noise PDF $p(\bfv_k)$ is Gaussian with mean $[0\ 0]^T$ and diagonal covariance matrix with $[1\ 0.1]$ on diagonal.

	Four versions of the grid-based point-mass filter (PMF) were implemented and tested, namely \textit{(i)} the standard PMF \citep{SiKraSo:06}, \textit{(ii)} the FFT based PMF \citep{DuMaSt:23}, \textit{(iii)} the proposed PMF in TT format (Algorithm 2), and \textit{(iv)} the proposed efficient PMF in TT format with FFT prediction (Algorithm 3). For the sake of completeness, \textit{(v)} the bootstrap particle filter (PF) \citep{SiKraSo:06} was compared as well.
	
	The performance of the efficient filters \textit{(ii)}--\textit{(v)} were compared against the standard PMF \textit{(i)} in terms of the relative difference of the root mean square error (RMSE). The results are given in Table \ref{tab:resAcc} which indicates negligible difference in performance of all grid based filters with a slight exception of the TT-based FFT implementation, where the difference is caused by a numerical instability of the interpolation step.
	
	\subsection{Curse of Dimensionality Reduction Illustration}
	
	Regarding the CC and storage requirement, the above mentioned PMFs were implemented for systems with varying state dimension $d$. The results can be seen in Fig. \ref{fig:timeStor}.
	
	The upper plot shows the behaviour of  Bytes plotted against dimension of the state needed to save the TPM and one PMD in the according format used by given filter. The curves of non-TT-based filters (standard PMF (STD) and FFT-based PMF) end at 3 dimensions as there was not enough RAM to save all data need for higher dimensional states, but they can be easily extrapolated. It can be seen that the TT based filters require similar storage which is, in the TT case, given mainly by the TPM matrix and in the TT FFT case,  by the TPM and PMD in the frequency domain. It is \textit{apparent} that the TT decomposition breaks the curse of dimensionality.
	
	The bottom plot shows an execution time dependence on $d$. Again the results for the standard PMF were not calculated for more than 3 dimensions due to a storage and CC. For the FFT filter, however, only the size of the RAM was a problem, the calculation is still fast. The TT FFT and TT PMFs again address the dimensionality issue. Since the interpolation algorithm used is not computationally optimised, an execution time without the interpolation time spent on interpolation named as ``TTFFT no interp'' is given for  completeness.

	It should be noted that the results for TT and FFT TT PMFs depend on the model, initial condition, and realisation of the \textit{random} greedy restricted cross interpolation algorithm. This is demonstrated by the dip in the TT storage complexity for $d=3$. 
	
	For $d \geq 3$ we have observed numerical instabilities in TT approximations as the greedy cross used does not guarantee positiveness of the underlying PDFs. The solution will be part of a future research.
		
		\begin{table}[t]
				\centering
				\begin{tabular}{lllll}
				 	& PMF \textit{(ii)} & PMF \textit{(iii)} & PMF \textit{(iv)}  & PF \textit{(v)} \\ 
								\hline 
					    RMSE$(x_k)$  & 0.0269 \%  & 0.0044   \%  & 1.7656  \%  &  0.40779  \% \\
						RMSE$(y_k)$ & 0.0309 \%  &  0.0355  \%  &  5.7051   \% & 0.44603  \% \\ 
						\hline 
					\end{tabular}
				\caption{Percentage accuracy difference.}
				\label{tab:resAcc}
			\end{table}
			
%
	
	\section{Concluding Remarks}
	The paper dealt with numerical solution to the BRRs using the grid-based filters. We have shown that the TT approximation is a powerful tool to break the curse of dimensionality of the filters. However, still the results should be viewed as the proof of concept. To turn the proposed filters into the practice in higher dimensions, a number of issues has to be resolved, such as oscillations in the tensor train approximations leading to negative values of the PMD, and CC scaling and oscillations of the interpolation algorithm.


\end{document}